\documentclass[journal]{IEEEtran}
\ifCLASSINFOpdf
\else
   \usepackage[dvips]{graphicx}
\fi
\usepackage{url}
\usepackage{graphicx}
\usepackage{amsmath}
\usepackage{verbatim}
\usepackage{amssymb}
\usepackage{algorithm}
\usepackage{algorithmic}
\usepackage{multirow}
\usepackage{tabularx}
\usepackage{array}
\usepackage{booktabs}
\usepackage{csquotes}
\usepackage[colorlinks, linkcolor=blue, citecolor=blue, urlcolor=cyan]{hyperref}
\usepackage{orcidlink}
\usepackage{threeparttable}

\begin{document}

\title{Efficient Streaming Voice Steganalysis in Challenging Detection Scenarios}

\author{Pengcheng Zhou\orcidlink{0009-0001-2659-1166}\textsuperscript{\dag}, Zhengyang Fang\orcidlink{0009-0006-6117-1059}\textsuperscript{\dag}, Zhongliang Yang\IEEEauthorrefmark{1}, Zhili Zhou, Linna Zhou

\thanks{
\textsuperscript{\dag} Equal Contribution. \IEEEauthorrefmark{1} Corresponding Author. 

Pengcheng Zhou is with the International School, Beijing University of Posts and Telecommunications, Beijing 100876, China.

Zhengyang Fang, Zhongliang Yang and Linna Zhou are with the School of Cyberspace Security, Beijing University of Posts and Telecommunications, Beijing 100876, China (e-mail: yangzl@bupt.edu.cn).

Zhili Zhou is with the School of Artificial Intelligence, Guangzhou University, Guangzhou 510006, China.
}
}

\markboth{Journal of \LaTeX\ Class Files, Vol. 14, No. 8, August 2015}
{Shell \MakeLowercase{\textit{et al.}}: Bare Demo of IEEEtran.cls for IEEE Journals}
\maketitle

\begin{abstract}
In recent years, there has been an increasing number of information hiding techniques based on network streaming media, focusing on how to covertly and efficiently embed secret information into real-time transmitted network media signals to achieve concealed communication. The misuse of these techniques can lead to significant security risks, such as the spread of malicious code, commands, and viruses. Current steganalysis methods for network voice streams face two major challenges: efficient detection under low embedding rates and short duration conditions. These challenges arise because, with low embedding rates (e.g., as low as 10\%) and short transmission durations (e.g., only 0.1 second), detection models struggle to acquire sufficiently rich sample features, making effective steganalysis difficult. To address these challenges, this paper introduces a Dual-View VoIP Steganalysis Framework (\textbf{DVSF}). The framework first randomly obfuscates parts of the native steganographic descriptors in VoIP stream segments, making the steganographic features of hard-to-detect samples more pronounced and easier to learn. It then captures fine-grained local features related to steganography, building on the global features of VoIP. Specially constructed VoIP segment triplets further adjust the feature distances within the model. Ultimately, this method effectively address the detection difficulty in VoIP. Extensive experiments demonstrate that our method significantly improves the accuracy of streaming voice steganalysis in these challenging detection scenarios, surpassing existing state-of-the-art methods and offering superior near-real-time performance.
\end{abstract}


\IEEEpeerreviewmaketitle

\section{Introduction}
\IEEEPARstart{C}{overt} systems are one of the three fundamental information security systems in the field of information security, alongside cryptographic systems and privacy systems \cite{shannon1949communication}. A notable characteristic of covert systems is their strong concealment of information. By embedding secret information into various ordinary carriers and transmitting it over public channels, they reduce the statistical distribution differences between carriers with embedded information and normal ones, thereby disguising the obvious communication activities. This method cleverly completes the process of hiding the information, thereby avoiding suspicion and attack \cite{simmons1984prisoners}.

Typically, a covert system can be modeled as the \enquote{prisoners' problem} \cite{simmons1984prisoners}. In this model, the sender needs to convey a secret message from a space of secret messages to the receiver. The sender selects any normal carrier from the public channel and uses a key to embed the secret message into the normal carrier. This transforms the original normal carrier into a steganographic carrier, and a large number of these steganographic carriers constitute the steganographic space. Subsequently, upon receiving the steganographic carrier, the receiver uses their own key to extract the secret message from it. To ensure the secrecy of the message, we typically require that the carrier remain completely identical in multidimensional features before and after steganography. However, in practice, this mapping often affects the probability distribution between the two. Overall, the primary goal of the sender and receiver is to successfully transmit information without arousing Eve's suspicion, so they need to minimize differences in the statistical distribution of carriers before and after steganography as much as possible. Conversely, Eve's task is to accurately determine whether a carrier contains secret information, so she needs to identify these differences as effectively as possible.

\begin{figure}[t]
\centering
\includegraphics[width=\columnwidth,height=0.2\textheight]{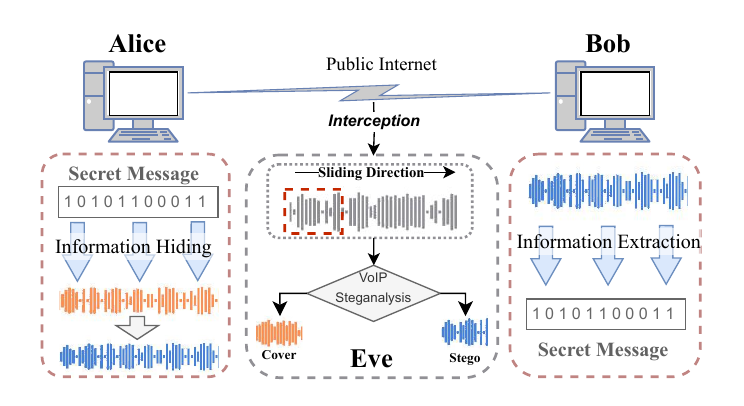}
\caption{In the VoIP steganalysis scenario, it is necessary to intercept the real-time voice streams transmitted over the network and efficiently analyze and determine whether they contain concealed information.}
\label{fig1}
\end{figure}

In today's social networks, various media formats, including images \cite{martin2023evolving, lu2021large, xu2022robust}, audio \cite{tian2014improving, chen2021learning, abood2022audio}, and text \cite{yang2021linguistic, gurunath2021novel, zhou2021linguistic, yang2020vae}, can serve as carriers for steganography. In recent years, with the widespread adoption and development of the Internet, communication via streaming media has significantly advanced. Network streaming media, as a real-time dynamic audio medium, contrasts with other static media carriers by having controllable transmission time and, theoretically, an infinite capacity for information hiding, which makes streaming media-based information hiding more challenging to detect. Voice over Internet Protocol (VoIP) is a widely adopted form of streaming media, directly connecting to IP network communications and offering lower communication costs. Common VoIP codecs include G.723.1 and G.729, both extensively using Linear Predictive Coding (LPC) due to its ability to achieve high compression ratios while maintaining satisfactory speech quality. During the encoding process, it is feasible to hide information within the payload (i.e., speech frames). Steganographic methods based on low bit-rate VoIP can generally be categorized into two types. The first type embeds secret information during the LPC encoding process \cite{wu2009lpc, xiao2008approach, tian2014improving}; for example, Xiao \emph{et al.} \cite{xiao2008approach} introduced the Complementary Neighbor Vertices (CNV) technique, which embeds secret information into Linear Spectrum Pair (LSP) parameters of the speech by improving codebook partitioning through graph theory in Quantization Index Modulation (QIM). The second type integrates secret information into pitch period prediction \cite{nishimura2009data, huang2012steganography, janicki2016pitch}, often referred to as Pitch Modulation Steganography (PMS). Huang \emph{et al.} \cite{huang2012steganography} demonstrated this by embedding secret information within the speech by modifying the Adaptive Codebook Delay (ACD) parameters. With the proliferation of the Internet and the widespread use of Internet telephony, these VoIP-based covert communication methods pose an increasing threat to cybersecurity. Therefore, studying VoIP steganalysis methods to achieve rapid and high-performance detection of near-real-time voice signal streams holds significant value and importance for safeguarding cyberspace and public safety.

The primary objective of VoIP steganalysis techniques is to detect whether steganographic information is embedded in real-time transmitted voice streams over the network by identifying statistical distribution differences between normal VoIP carriers and steganographic VoIP carriers, as illustrated in Figure 1. Compared to steganalysis methods for static carriers, those for streaming media like VoIP are more complex and demanding. Since VoIP signals are transmitted in real-time online, detection algorithms must be efficient enough to handle this real-time nature. This imposes two main requirements on VoIP steganalysis methods: firstly, the steganalysis model must efficiently detect very short voice signals, such as those with a duration of 0.1 second, to ensure that steganographic communications can be identified shortly after they begin, allowing any suspicious communication behaviors to be terminated promptly. Secondly, the steganalysis model needs to maintain high accuracy to avoid unnecessary false positives and false negatives. This is particularly crucial in detection scenarios with low embedding rates (e.g., 10\%) or short segment lengths (e.g., 0.1 second). In these scenarios, VoIP segments contain sparse steganographic features, making them challenging to detect.

Although VoIP steganalysis is challenging, many researchers have begun to explore techniques in this area. Initially, VoIP steganalysis methods relied on handcrafted techniques or model-based learning approaches, typically analyzing the statistical characteristics of carriers to determine whether a VoIP speech segment contains hidden information. Examples include Mel-cepstral steganalysis \cite{kraetzer2007mel} and codeword correlation \cite{li2017steganalysis, li2012detection}, and so on \cite{huang2011detection, dittmann2005steganography}. These methods often struggle to balance detection efficiency and accuracy, with some requiring significant time for feature extraction and analysis to achieve relatively high detection precision, rendering them unsuitable for near-real-time detection \cite{li2017steganalysis}. In recent years, the rapid advancement of artificial intelligence (AI) technologies has significantly enhanced the accuracy and automation of steganalysis. These modern approaches utilize feature extraction, classification, and deep neural networks, including convolutional neural networks (CNN) \cite{yang2022real, li2014detection, 10107475}, recurrent neural networks (RNN) \cite{lin2018rnn}, attention mechanisms \cite{wang2021fast, zhang2024tenet}, and hybrid neural networks \cite{yang2019fast, hu2021detection, li2017steganalysis, yang2019steganalysis} to analyze and detect steganographic information in VoIP. Most of these works focus on the detection performance of models on VoIP segments with pronounced steganographic features. However, they often struggle to produce satisfactory results in practical scenarios where steganographic features are subtle, such as in low embedding rates or short-duration VoIP segment steganalysis.

To further advance VoIP steganalysis technology and enhance its practical value, we focus on addressing two major challenges in the field: the difficulty of detection in low embedding rate or short-duration scenarios and the need for near-real-time detection. In this paper, we design and propose a novel Dual-View VoIP Steganalysis Framework (\textbf{DVSF}), where we develop a Hybrid Attention Model (\textbf{HAM}) based on convolution and attention mechanisms focused on the fusion analysis of local and global steganographic features in VoIP. During the data preprocessing phase, we employ the CutMix technique to mix and diffuse the steganographic descriptors of more prominent VoIP segments with those of hard-to-detect VoIP segments. This is done to adjust the feature quantities carried by the hard-to-detect VoIP segments, which helps improve the model's focus on these challenging VoIP segments. This framework leverages the advantages of contrastive learning in feature representation learning, adjusting the feature distance between normal and steganographic VoIP segments within the \textbf{HAM}'s feature space to a state that is easily linearly separable, thereby enhancing the model's steganalysis performance. This process is based on specially constructed VoIP segment triplets. Finally, from an application perspective, extensive experimental results demonstrate that our proposed method significantly outperforms existing methods in challenging VoIP steganography detection scenarios, better adapting to more realistic and complex data situations, proving the reliability of our approach. The relevant experimental results and analyses show significant improvements in steganography detection capability.

The rest of the paper is organized as follows: Section II introduces related work on VoIP steganography and steganalysis. Section III details the proposed framework. The subsequent Section IV presents our validation experiments, including the experimental setup and results. Finally, we draw conclusions in Section V.

\section{Related Works}
\subsection{Steganography for Low Bit-Rate VoIP}
VoIP, as a form of streaming media, is widely used in communication networks, utilizing existing IP networks to transmit voice data with low bandwidth requirements, thereby significantly reducing the technical costs of services like telephony. Due to the capability of Linear Predictive Coding (LPC) to achieve high compression ratios and satisfactory voice quality, low bit-rate speech codecs based on LPC, such as G.723.1 and G.729, are extensively deployed in VoIP. Steganographic methods based on low bit-rate VoIP voice streams generally fall into two categories. The first category hides secret information during the LPC encoding process \cite{wu2009lpc, xiao2008approach, tian2014improving}. A representative method is CNV-QIM proposed by Xiao \emph{et al.} \cite{xiao2008approach}, which uses carefully designed codebook partitioning and is based on Quantization Index Modulation (QIM). This optimizes the embedding process of information into speech parameters, reducing audio quality degradation. The second category integrates information hiding into pitch period prediction \cite{nishimura2009data, huang2012steganography, janicki2016pitch}. A representative method is PMS proposed by Huang \emph{et al.} \cite{huang2012steganography}, which achieves information embedding by adjusting the pitch period search range, without significantly impacting audio quality and latency.

\begin{figure*}[t]
\centering
\includegraphics[width=\textwidth]{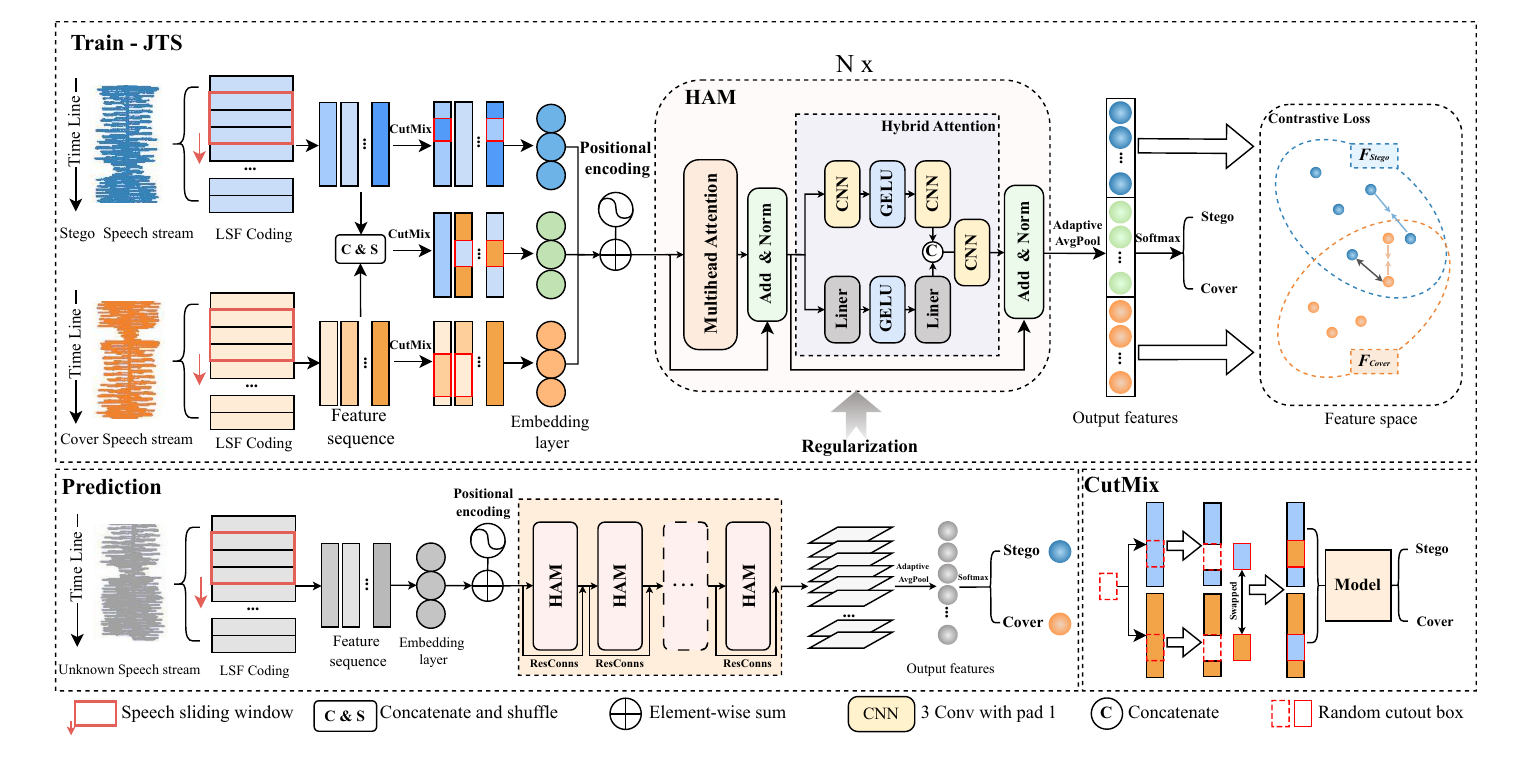}
\caption{An overview of the proposed framework DVSF, which consists of three integral components: the CutMix module, the Hybrid Attention Model (\textbf{HAM}), and the Joint Training Strategy (\textbf{JTS}), encompassing both the training and prediction processes. The VoIP segment triplet features are solely the outputs of the model during the training phase and are not required in the prediction phase.}
\label{fig2}
\end{figure*}


\subsection{VoIP Steganalysis}
VoIP steganalysis aims to detect secret information within VoIP streams to prevent the illegal use of steganography. From the perspective of adaptability to steganography, steganalysis methods are typically divided into two categories: general methods \cite{kraetzer2007mel, pelaez2009using} and targeted methods \cite{li2012detection, li2017steganalysis, yang2018steganalysis, ren2016amr, jia2015detection, hu2021detection, wang2021fast}. General methods often rely on manually constructed or model self-learning approaches to analyze the statistical distribution characteristics of carriers. Kraetzer \emph{et al.} \cite{kraetzer2007mel} utilized a method based on Mel-cepstral analysis for detecting embedded hidden messages and combined it with traditional audio steganalysis features to improve detection performance. Pelaez \emph{et al.} \cite{pelaez2009using} systematically analyzed steganography in VoIP using a misuse pattern to enhance the ability to detect embedded hidden messages in IP telephony. The weakness of these methods is that they do not include feature analysis specific to particular VoIP steganographic carriers, leading to lower detection accuracy for specific VoIP steganography.

The second category of VoIP steganalysis methods aims to detect secret information hidden using specific steganographic algorithms. To detect CNV-QIM steganography \cite{xiao2008approach}, Li \emph{et al.} \cite{li2017steganalysis} developed a model based on Quantization Codeword Correlation Network (SS-QCCN) for detection. Yang \emph{et al.} \cite{yang2018steganalysis} employed a steganalysis approach based on a Codeword Bayesian Network (CBN), utilizing the probability distribution of codewords and sensitive transitions. In 2023, Wei \emph{et al.} \cite{2023Frame} proposed a frame-level steganalysis method based on a neural network classification framework of multidimensional codeword correlation, achieving steganography detection. In 2024, Zhang \emph{et al.} \cite{10574375} proposed the LStegT model, which significantly reduces computational resource consumption by combining one-dimensional depthwise separable convolutions with a Transformer architecture. For PMS \cite{huang2012steganography} detection, Li \emph{et al.} \cite{li2014detection} developed a Codebook Correlation Network (CCN) that uses conditional probability to quantify the correlation between adaptive codebook coefficients, enabling efficient detection of this steganographic method. Ren \emph{et al.} \cite{ren2016amr} achieved efficient detection of steganographic audio by calculating the second-order difference matrix features of the audio and employing a calibration method. These methods all focus solely on the steganalysis techniques themselves and lack consideration of specific application scenarios. Additionally, in 2021, Hu \emph{et al.} \cite{hu2021detection} constructed a Steganalysis Feature Fusion Network (SFFN) using RNN and CNN to achieve effective steganalysis. In the same year, Wang \emph{et al.} \cite{wang2021fast} implemented a more efficient steganalysis using a neural network (KFEF) based on a multi-head attention mechanism and completed a preliminary assessment of the steganographic locations. These two methods differ from others in that they consider multiple steganographic techniques, yet they still face the aforementioned issues.

In summary, existing methods generally focus on the overall performance of VoIP steganalysis across various scenarios but lack specific discussion on more valuable application contexts. In real-world applications, these methods still face challenges such as the difficulty of detecting low embedding rates or short-duration samples. The near-real-time performance of detection methods is also a consideration.

\section{The Proposed Method}
The objective of our method is to efficiently identify the modified frame contents within VoIP segments that have been altered through steganography. Figure 2 provides an overview of our method's architecture (\textbf{DVSF}). In this process, we utilize the preprocessed VoIP steganographic descriptors \( d_o \in \mathbb{R}^{D \times T} \) and apply CutMix to obfuscate them, resulting in the obfuscated steganographic descriptors \( d_c \), where \( D \) represents the dimensionality of the steganographic descriptors and \( T \) denotes the number of voice frames, determined by the segment length. The steganographic descriptors \( d_c \) serves as the input to our model (\textbf{HAM}). Subsequently, the dual-view feature extractor effectively addresses both global and local features, fitting the authentic VoIP steganographic characteristics in high-dimensional space. During the model training phase, the training objective is denoted as \( \mathcal{L}_{joint} \), which comprises two components: the supervised contrastive learning loss function \( \mathcal{L}_{scl} \) and the cross-entropy loss function \( \mathcal{L}_{ce} \).

\subsection{VoIP Preprocessing}
For a continuous VoIP stream, the prevailing approach is to eliminate the continuity of the analysis carrier, extracting the non-continuous key features affected by steganography from the VoIP stream. Each VoIP segment can be regarded as a collection of voice frames \( f = \{f_1, f_2, \ldots, f_T\} \), where \( T \) denotes the number of sample voice frames, and each voice frame serves as the smallest unit for extracting critical steganographic features. Here, we detail the data processing using the G.729 codec as an example. In the G.729 codec, each 10 ms audio frame consists of two subframes, encoded into multiple parameters totaling 80 bits. For each voice frame \( f_t \in f, 1 \leq t \leq T \), the CNV-QIM steganography embeds secret information during the vector quantization of Linear Predictive Coding (LPC) coefficients, while the PMS steganography incorporates secret information during pitch analysis by adjusting the pitch delay to achieve information concealment, they are all within the LSF coding.

Thus, we extract LSP codewords from each voice frame to serve as the steganographic descriptors for CNV-QIM. Following the recommendations of Hu \emph{et al.} \cite{hu2021detection}, and to mitigate the adverse effects of encoding, we decode the Adaptive Codebook Delay (ACD) codewords into pitch delay, which we then utilize as the steganographic descriptors for PMS. Ultimately, for each continuous VoIP segment, we obtain non-continuous LSP codewords \( L = \{l_1, l_2, \ldots, l_T\} \) with a shape of \( (3, T) \) and pitch delays \( P = \{p_1, p_2, \ldots, p_T\} \) with a shape of \( (4, T) \) to serve as the input data for the subsequent stages of our method.

\subsection{CutMix for VoIP segments}
Low embedding rate or short-duration VoIP segments are challenging to detect due to their less pronounced steganographic features. To enhance the feature content of these hard-to-detect samples, we employ the CutMix \cite{2019CutMix} technique to blend the features of the samples. Specifically, by exchanging portions of the VoIP segment steganographic descriptors \( d_o = L \) or \( P \), CutMix introduces diversity at the feature level, allowing the model to encounter more prominent and diverse combinations of data during the training process. This operation effectively enhances the model's learning capability for hard-to-detect samples, as it compels the model to achieve a deeper understanding and generalization at the feature level. Consequently, when faced with more challenging detection scenarios, the model demonstrates improved robustness and accuracy. We define the more pronounced sample steganographic descriptors as \( (d_o^p) \) and the hard-to-detect sample steganographic descriptors as \( (d_o^h) \), with their corresponding labels denoted as \( y_o^p \) and \( y_o^h \). The generated samples are represented as \( (d_c, y_c) \), and this process can be expressed as follows:
\begin{equation}
\begin{aligned}
d_c &= \mathbf{M} \odot d_o^p + (\mathbf{1 - M}) \odot d_o^h, \\
y_c &= \lambda y_o^p + (1 - \lambda)y_o^h,
\end{aligned}
\end{equation}
where \(\odot\) denotes the element-wise multiplication. The value of \(\lambda\) is sampled from the beta distribution \(\mathbf{Beta}(\alpha, \alpha)\). To ensure the effectiveness of CutMix in our experiments, we set \(\alpha\) to 0.6, which allows for a larger exchange of feature values. The binary mask \(\mathbf{M} \in \{0,1\}^{D \times T}\) consists exclusively of 0s and 1s, and it is derived using the following method:
\begin{equation}
\mathbf{M(i,j)}=
\begin{cases}
0, & (i,j) \in \mathbf{C} \\
1, & (i,j) \notin \mathbf{C}
\end{cases}
, \mathbf{C} = (C_x, C_y, C_{depth}, C_{hight}),
\end{equation}
where \(\mathbf{C}\) represents the CutMix region of the VoIP segment steganographic descriptors \( d_o \), with the center coordinate defined as \((C_x, C_y)\), where \(C_x \sim \text{Uniform}(0, D)\) and \(C_y \sim \text{Uniform}(0, T)\). The dimensions of the region are denoted by \((C_{depth}, C_{hight})\), where \(C_{depth} = D \times \sqrt{1 - \lambda}\) and \(C_{hight} = T \times \sqrt{1 - \lambda}\). Therefore, the CutMix region for VoIP segments is a matrix of random position and random size. 

\subsection{Dual-View Feature Extractor}
We designed and implemented a Hybrid Attention Model (\textbf{HAM}) for the fusion analysis of global and local steganographic features. Its primary aim is to leverage the global perceptive capabilities of the attention mechanism, which is not constrained by sequence length, while extending this with a hybrid attention network formed by Convolutional Neural Network (CNN) and GELU Feedforward Neural Network (GFFN). This architecture enables the model to comprehend the overall structure of VoIP steganographic features while simultaneously capturing subtle local changes, thereby enhancing its adaptability to the characteristics of hard-to-detect samples. We refer to this as the dual-view feature extractor.

In the \textbf{HAM} block, we take the VoIP steganographic descriptors \( d_c \) and pass them through an embedding layer and a sine-cosine position encoding layer to obtain the high-dimensional features \( F_{in} \in \mathbb{R}^{D \times B \times M} \) as the input to the model. Here, \( B \) is three times the training batch size, which relates to our specific training approach, and \( M \) is the model dimension. We will now provide a detailed description of the feature extraction steps:

We first utilize a multi-head self-attention mechanism to extract the long-range features of the VoIP segments, allowing us to preliminarily capture the global VoIP steganographic feature information \( F_{global} \in \mathbb{R}^{D \times B \times M} \). This process can be expressed as follows:
\begin{equation}
F_{global} = LN(F_{in} + Dropout(MSA(F_{in}))),
\end{equation}
where \( MSA(\cdot) \) denotes the multi-head self-attention computation module. Subsequently, we designed two parallel subnetwork branches comprising a Convolutional Neural Network (CNN) and a GELU Feedforward Neural Network (GFFN) to analyze the more concealed local feature information from the existing global VoIP features \( F_{global} \). This approach allows the model to place greater emphasis on features that are less apparent in hard-to-detect samples. The architecture of the subnetwork can be represented as follows:
\begin{equation}
\begin{aligned}
CNN(\cdot) = &conv_2(GELU(conv_1(\cdot))), \\
GFFN(\cdot) = l&inear_2(GELU(linear_1(\cdot))).
\end{aligned}
\end{equation}

We concatenate the feature information extracted from the two subnetworks and use a CNN for feature fusion, thereby forming our hybrid attention network \( HA(\cdot) \). This results in the integrated local features regarding VoIP steganography, represented as \( F_{out} \in \mathbb{R}^{B \times M \times D} \). The process can be expressed as follows:
\begin{equation}
\begin{aligned}
HA(\cdot) &= conv_3(cat(CNN(\cdot),GFFN(\cdot))), \\
F_{out} = L&N(F_{global} + Dropout(HA(F_{global}))).
\end{aligned}
\end{equation}

We stack multiple \textbf{HAM} blocks together to form the dual-view feature extractor, which can be expressed as follows:
\begin{equation}
H^n = HAM_n(H^{n-1}), 1 \leq n \le N,
\end{equation}
where \( H^n \) represents the output of the \( n \)-th \textbf{HAM} block, and \( N \) denotes the total number of \textbf{HAM} blocks.

\subsection{Joint Training Strategy}
We designed our training strategy primarily utilizing a supervised contrastive learning objective, complemented by a cross-entropy loss objective. To address the detection challenges posed by the less pronounced features of VoIP segments, we cleverly constructed cyclic sample triplets \( X^{t} \in \mathbb{R}^{N \times B \times M \times D} \) suitable for supervised contrastive learning, where \( N \) is the number of triplet batches, \( B \) is simply three times the training batch size without any specific meaning, as determined by the triplet construction method. The method for constructing VoIP segment triplets and the notation definitions can be found in Algorithm 1. The sample triplets are constructed by cyclically using positive samples, specifically, the positive VoIP segment pair $(X^{+'}_{is}, X^{+'}_{ir}), r = (s \ mod \ B_{tri}) + 1$ in the last triplet of each batch consists of the last and the first VoIP segments from the original positive sample batch, \(B_{tri}\) denotes the batch size of the VoIP segment triplet. In the hyperspherical feature space of the model, constructing VoIP segment triplets for training aims to increase the similarity of features among similar VoIP segments by forming closer positive sample pairs. It also seeks to reduce the expected distance between hard-to-detect VoIP segments to optimize alignment and uniformity \cite{wang2020understanding}, resulting in a more uniform distribution of features for various VoIP segments. This approach facilitates linear separability between different VoIP segments within the model's feature space, as shown in Figure 3.

\begin{figure}[h]
\centering
\includegraphics[width=\columnwidth]{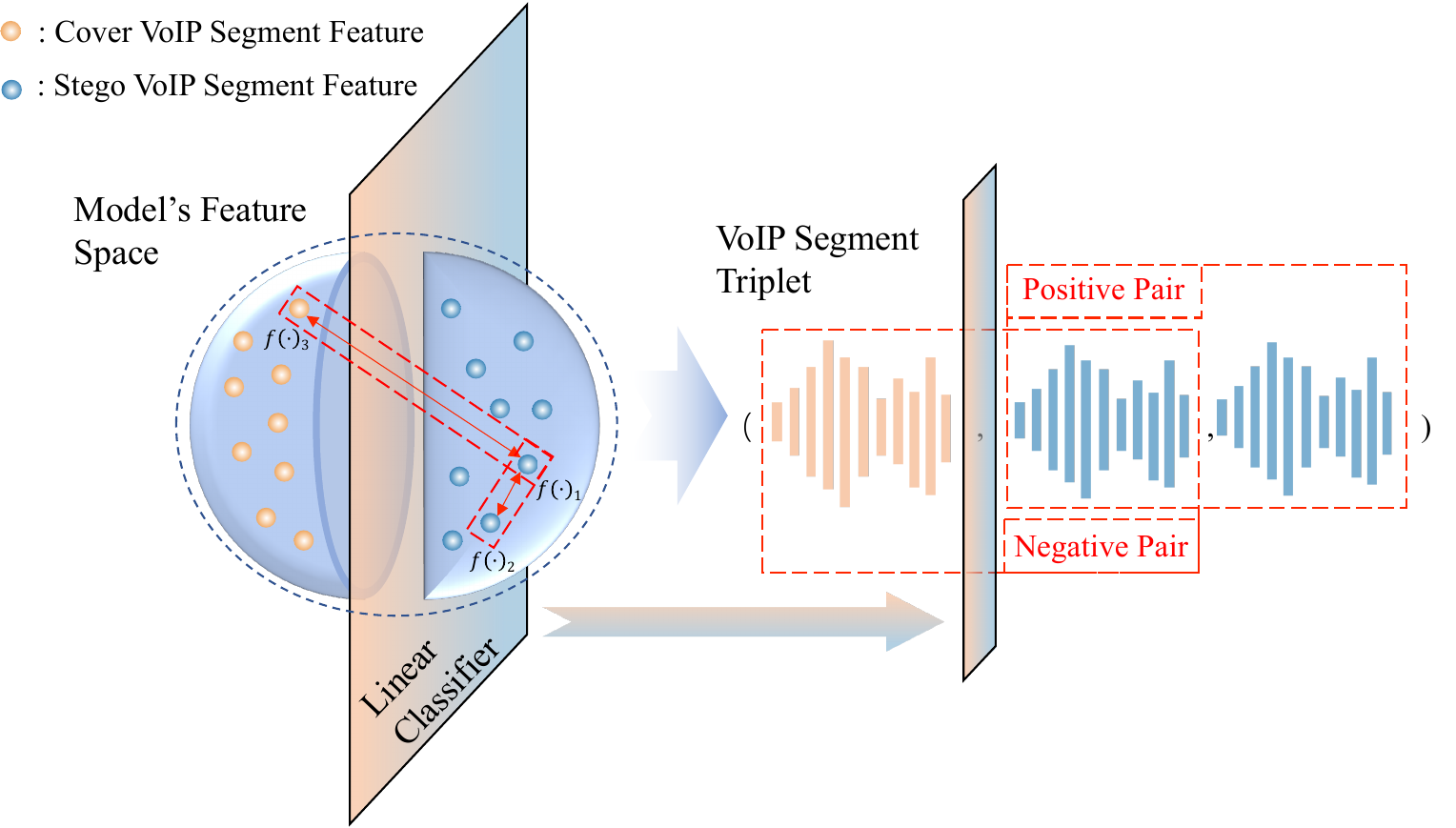}
\caption{The Joint Training Strategy promotes the alignment and uniformity of feature distribution in the model feature space by influencing the feature distances within positive and negative VoIP segment pairs, thereby making the feature space linearly separable.}
\label{fig:Linear_Classification}
\end{figure}

Noise during the model training process is unavoidable, and learning from hard-to-detect VoIP segments, which have less feature content, is often more susceptible to noise. Thus, it is essential to reduce the parameter values to avoid overly complex models fitting strong features from easily detectable VoIP segments, allowing the model to better focus on the features of hard-to-detect VoIP segments and improving its robustness against noise when fitting their features. This process can be expressed as follows:
\begin{equation}
\theta_{HAM} \gets \theta_{HAM} - \omega\theta_{HAM},
\end{equation}
where \(\omega\) denotes the parameter decay coefficient.

Next, we introduce BatchNorm layers to enrich the information exchange between the features of different VoIP segments, enhancing the model's training generalization. After employing average pooling layers to downsample the high-dimensional VoIP features, we establish the following supervised contrastive learning training objective:
\begin{equation}
  \mathcal{L}_{scl} = -\log \frac{e^{\text{sim}(h_i, h^+_i)/\tau}}{\sum_{j=1}^{B_{tri}} (e^{\text{sim}(h^+_i, h^+_j)/\tau} + e^{\text{sim}(h^+_i, h^-_j)/\tau})},
\end{equation}
where \(\tau\) is a temperature hyperparameter used to adjust the feature distances between normal and steganographic VoIP segments. Here, \(h_i\) and \(h_j\) represent the high-dimensional features of the VoIP segments.

\begin{algorithm}[t]
    \caption{algorithm for constructing VoIP segment triplets}
    \label{JTS}
    \renewcommand{\algorithmicrequire}{\textbf{Input:}}
    \renewcommand{\algorithmicensure}{\textbf{Output:}}
    \begin{algorithmic}[1]
        \REQUIRE \(\mathbf{X^+} = \{X^+_1, X^+_2, \ldots, X^+_p\}\) represents the collection of positive VoIP segment batches, where \(X^+_i\) denotes the i-th batch of positive VoIP segments. \(\mathbf{X^-} = \{X^-_1, X^-_2, \ldots, X^-_n\}\) denotes the collection of negative VoIP segment batches, where \(X^-_j\) is the j-th batch of negative VoIP segments
        \ENSURE \(\mathbf{X^t} = \{X^t_1,X^t_2,\ldots,X^t_m\}\), where \(X^t_k\) is the k-th batch of triplet VoIP segments
        
        \STATE $j \gets 1$
        \FOR{$i = 1$ to $p$}

            \STATE $k \gets i$

            \STATE $min\_size \gets \min(\text{size}(X^+_{i}), \text{size}(X^-_{j}))$
            
            \STATE $X^{+'}_{i} \gets (X^+_{i}[:min\_size])$
            \STATE $X^{-'}_{j} \gets (X^-_{j}[:min\_size])$
            
            \FOR{$s = 1$ to $min\_size$}

                \STATE $r \gets (s \mod min\_size) + 1$
                \STATE $X^{t}_{ks} \gets (X^{+'}_{is}, X^{+'}_{ir}, X^{-'}_{js})$

            \ENDFOR
            
            \STATE $j \gets (j \mod n) + 1$
        \ENDFOR
    \end{algorithmic}
\end{algorithm}

We further reduce the dimensionality of the VoIP segment features through a fully connected layer and use the Softmax function to compute the probability results for each VoIP segment being classified as either normal or steganographic. We establish the following cross-entropy loss training objective to further encourage model convergence and enhance its steganalysis capabilities:
\begin{equation}
\mathcal{L}_{ce} = -(y \cdot \log(\hat{y}) + (1 - y) \cdot \log(1 - \hat{y})).
\end{equation}

\begin{table*}[t]
\centering
\caption{For VoIP segment dataset $D_e$, the detection performance (accuracy) of each method at different embedding rates.}
\label{table1}
\newcolumntype{Y}{>{\centering\arraybackslash}X}
\setlength{\extrarowheight}{2.5pt}
\begin{tabularx}{\textwidth}{YYYYYYYYYYYY}

\midrule[1.5pt]
\multicolumn{1}{l}{\textbf{Steganalysis}} & \multicolumn{1}{l|}{\textbf{Steganography}} & \multicolumn{10}{c}{\textbf{Embedding Rate (\%)}}\\

\multicolumn{1}{l}{\textbf{Method}} & \multicolumn{1}{l|}{\textbf{Method}} & \multicolumn{1}{Y}{10} & \multicolumn{1}{Y}{20} & \multicolumn{1}{Y}{30} & \multicolumn{1}{Y}{40} & \multicolumn{1}{Y}{50} & \multicolumn{1}{Y}{60} & \multicolumn{1}{Y}{70} & \multicolumn{1}{Y}{80} & \multicolumn{1}{Y}{90} & \multicolumn{1}{Y}{100}\\
\hline
\multicolumn{1}{c}{\multirow{2}{*}{CCN \cite{li2014detection}}} & \multicolumn{1}{c|}{CNV-QIM \cite{xiao2008approach}} & \multicolumn{1}{c}{/} & \multicolumn{1}{c}{/} & \multicolumn{1}{c}{/} & \multicolumn{1}{c}{/} & \multicolumn{1}{c}{/} & \multicolumn{1}{c}{/} & \multicolumn{1}{c}{/} & \multicolumn{1}{c}{/} & \multicolumn{1}{c}{/} & \multicolumn{1}{c}{/}\\

\multicolumn{1}{c}{} & \multicolumn{1}{c|}{PMS \cite{huang2012steganography}} & \multicolumn{1}{c}{53.75} & \multicolumn{1}{c}{54.85} & \multicolumn{1}{c}{58.30} & \multicolumn{1}{c}{65.15} & \multicolumn{1}{c}{69.00} & \multicolumn{1}{c}{74.65} & \multicolumn{1}{c}{80.05} & \multicolumn{1}{c}{84.90} & \multicolumn{1}{c}{87.70} & \multicolumn{1}{c}{90.20}\\
\hline
\multicolumn{1}{c}{\multirow{2}{*}{SS-QCCN \cite{li2017steganalysis}}} & \multicolumn{1}{c|}{CNV-QIM \cite{xiao2008approach}} & \multicolumn{1}{c}{55.00} & \multicolumn{1}{c}{57.85} & \multicolumn{1}{c}{67.77} & \multicolumn{1}{c}{80.86} & \multicolumn{1}{c}{85.90} & \multicolumn{1}{c}{90.90} & \multicolumn{1}{c}{94.25} & \multicolumn{1}{c}{96.20} & \multicolumn{1}{c}{97.80} & \multicolumn{1}{c}{98.65}\\

\multicolumn{1}{c}{} & \multicolumn{1}{c|}{PMS \cite{huang2012steganography}} & \multicolumn{1}{c}{/} & \multicolumn{1}{c}{/} & \multicolumn{1}{c}{/} & \multicolumn{1}{c}{/} & \multicolumn{1}{c}{/} & \multicolumn{1}{c}{/} & \multicolumn{1}{c}{/} & \multicolumn{1}{c}{/} & \multicolumn{1}{c}{/} & \multicolumn{1}{c}{/}\\
\hline
\multicolumn{1}{c}{\multirow{2}{*}{LStegT \cite{10574375}$^\dagger$}} & \multicolumn{1}{c|}{CNV-QIM \cite{xiao2008approach}} & \multicolumn{1}{c}{69.23} & \multicolumn{1}{c}{83.17} & \multicolumn{1}{c}{91.13} & \multicolumn{1}{c}{95.51} & \multicolumn{1}{c}{97.80} & \multicolumn{1}{c}{98.55} & \multicolumn{1}{c}{99.46} & \multicolumn{1}{c}{99.82} & \multicolumn{1}{c}{99.85} & \multicolumn{1}{c}{99.98}\\

\multicolumn{1}{c}{} & \multicolumn{1}{c|}{PMS \cite{huang2012steganography}} & \multicolumn{1}{c}{/} & \multicolumn{1}{c}{/} & \multicolumn{1}{c}{/} & \multicolumn{1}{c}{/} & \multicolumn{1}{c}{/} & \multicolumn{1}{c}{/} & \multicolumn{1}{c}{/} & \multicolumn{1}{c}{/} & \multicolumn{1}{c}{/} & \multicolumn{1}{c}{/}\\
\hline
\multicolumn{1}{c}{\multirow{2}{*}{FS-MDP \cite{2023Frame}$^\dagger$}} & \multicolumn{1}{c|}{CNV-QIM \cite{xiao2008approach}} & \multicolumn{1}{c}{54.78} & \multicolumn{1}{c}{59.11} & \multicolumn{1}{c}{65.00} & \multicolumn{1}{c}{73.12} & \multicolumn{1}{c}{74.89} & \multicolumn{1}{c}{76.44} & \multicolumn{1}{c}{81.12} & \multicolumn{1}{c}{83.31} & \multicolumn{1}{c}{86.66} & \multicolumn{1}{c}{88.69}\\

\multicolumn{1}{c}{} & \multicolumn{1}{c|}{PMS \cite{huang2012steganography}} & \multicolumn{1}{c}{/} & \multicolumn{1}{c}{/} & \multicolumn{1}{c}{/} & \multicolumn{1}{c}{/} & \multicolumn{1}{c}{/} & \multicolumn{1}{c}{/} & \multicolumn{1}{c}{/} & \multicolumn{1}{c}{/} & \multicolumn{1}{c}{/} & \multicolumn{1}{c}{/}\\
\hline
\multicolumn{1}{c}{\multirow{2}{*}{SFFN \cite{hu2021detection}}} & \multicolumn{1}{c|}{CNV-QIM \cite{xiao2008approach}} & \multicolumn{1}{c}{54.28} & \multicolumn{1}{c}{54.62} & \multicolumn{1}{c}{71.23} & \multicolumn{1}{c}{75.08} & \multicolumn{1}{c}{82.19} & \multicolumn{1}{c}{84.90} & \multicolumn{1}{c}{87.43} & \multicolumn{1}{c}{89.66} & \multicolumn{1}{c}{91.86} & \multicolumn{1}{c}{93.50}\\

\multicolumn{1}{c}{} & \multicolumn{1}{c|}{PMS \cite{huang2012steganography}} & \multicolumn{1}{c}{51.17} & \multicolumn{1}{c}{53.76} & \multicolumn{1}{c}{61.52} & \multicolumn{1}{c}{63.94} & \multicolumn{1}{c}{69.74} & \multicolumn{1}{c}{72.80} & \multicolumn{1}{c}{76.11} & \multicolumn{1}{c}{79.07} & \multicolumn{1}{c}{82.04} & \multicolumn{1}{c}{83.93}\\
\hline
\multicolumn{1}{c}{\multirow{2}{*}{KFEF \cite{wang2021fast}}} & \multicolumn{1}{c|}{CNV-QIM \cite{xiao2008approach}} & \multicolumn{1}{c}{62.71} & \multicolumn{1}{c}{74.27} & \multicolumn{1}{c}{83.01} & \multicolumn{1}{c}{88.58} & \multicolumn{1}{c}{92.35} & \multicolumn{1}{c}{94.80} & \multicolumn{1}{c}{96.47} & \multicolumn{1}{c}{97.59} & \multicolumn{1}{c}{98.42} & \multicolumn{1}{c}{98.90}\\

\multicolumn{1}{c}{} & \multicolumn{1}{c|}{PMS \cite{huang2012steganography}} & \multicolumn{1}{c}{53.50} & \multicolumn{1}{c}{57.80} & \multicolumn{1}{c}{63.02} & \multicolumn{1}{c}{67.16} & \multicolumn{1}{c}{71.28} & \multicolumn{1}{c}{74.85} & \multicolumn{1}{c}{77.70} & \multicolumn{1}{c}{81.09} & \multicolumn{1}{c}{83.68} & \multicolumn{1}{c}{85.76}\\
\hline
\hline
\multicolumn{1}{c}{\multirow{2}{*}{DVSF (ours)}} & \multicolumn{1}{c|}{CNV-QIM \cite{xiao2008approach}} & \multicolumn{1}{c}{\textbf{79.03}} & \multicolumn{1}{c}{\textbf{91.75}} & \multicolumn{1}{c}{\textbf{96.88}} & \multicolumn{1}{c}{\textbf{98.72}} & \multicolumn{1}{c}{\textbf{99.27}} & \multicolumn{1}{c}{\textbf{99.89}} & \multicolumn{1}{c}{\textbf{99.95}} & \multicolumn{1}{c}{\textbf{99.99}} & \multicolumn{1}{c}{\textbf{99.99}} & \multicolumn{1}{c}{\textbf{99.99}}\\

\multicolumn{1}{c}{} & \multicolumn{1}{c|}{PMS \cite{huang2012steganography}} & \multicolumn{1}{c}{\textbf{62.45}} & \multicolumn{1}{c}{\textbf{70.85}} & \multicolumn{1}{c}{\textbf{78.90}} & \multicolumn{1}{c}{\textbf{84.15}} & \multicolumn{1}{c}{\textbf{88.66}} & \multicolumn{1}{c}{\textbf{91.86}} & \multicolumn{1}{c}{\textbf{94.32}} & \multicolumn{1}{c}{\textbf{96.64}} & \multicolumn{1}{c}{\textbf{97.80}} & \multicolumn{1}{c}{\textbf{98.31}}\\

\midrule[1.5pt]

\end{tabularx}
\begin{threeparttable}
 \begin{tablenotes}
        \scriptsize
        \item[$\dagger$] We implement the methods on our datasets.
\end{tablenotes}
\end{threeparttable}
\end{table*}

\begin{table*}[t]
\centering
\caption{For VoIP segment dataset $D_s$, the detection performance (accuracy) of each method at different segment length.}
\label{table2}
\newcolumntype{Y}{>{\centering\arraybackslash}X}
\setlength{\extrarowheight}{2.5pt}
\begin{tabularx}{\textwidth}{YYYYYYYYYYYY}

\midrule[1.5pt]
\multicolumn{1}{l}{\textbf{Steganalysis}} & \multicolumn{1}{l|}{\textbf{Steganography}} & \multicolumn{10}{c}{\textbf{Segment Length (s)}}\\

\multicolumn{1}{l}{\textbf{Method}} & \multicolumn{1}{l|}{\textbf{Method}} & \multicolumn{1}{Y}{0.1} & \multicolumn{1}{Y}{0.2} & \multicolumn{1}{Y}{0.3} & \multicolumn{1}{Y}{0.4} & \multicolumn{1}{Y}{0.5} & \multicolumn{1}{Y}{0.6} & \multicolumn{1}{Y}{0.7} & \multicolumn{1}{Y}{0.8} & \multicolumn{1}{Y}{0.9} & \multicolumn{1}{Y}{1.0}\\
\hline
\multicolumn{1}{c}{\multirow{2}{*}{CCN \cite{li2014detection}}} & \multicolumn{1}{c|}{CNV-QIM \cite{xiao2008approach}} & \multicolumn{1}{c}{/} & \multicolumn{1}{c}{/} & \multicolumn{1}{c}{/} & \multicolumn{1}{c}{/} & \multicolumn{1}{c}{/} & \multicolumn{1}{c}{/} & \multicolumn{1}{c}{/} & \multicolumn{1}{c}{/} & \multicolumn{1}{c}{/} & \multicolumn{1}{c}{/}\\

\multicolumn{1}{c}{} & \multicolumn{1}{c|}{PMS \cite{huang2012steganography}} & \multicolumn{1}{c}{65.25} & \multicolumn{1}{c}{72.25} & \multicolumn{1}{c}{77.20} & \multicolumn{1}{c}{79.50} & \multicolumn{1}{c}{82.75} & \multicolumn{1}{c}{85.80} & \multicolumn{1}{c}{86.90} & \multicolumn{1}{c}{89.10} & \multicolumn{1}{c}{90.05} & \multicolumn{1}{c}{90.55}\\
\hline
\multicolumn{1}{c}{\multirow{2}{*}{SS-QCCN \cite{li2017steganalysis}}} & \multicolumn{1}{c|}{CNV-QIM \cite{xiao2008approach}} & \multicolumn{1}{c}{79.60} & \multicolumn{1}{c}{88.70} & \multicolumn{1}{c}{93.10} & \multicolumn{1}{c}{94.20} & \multicolumn{1}{c}{95.50} & \multicolumn{1}{c}{96.80} & \multicolumn{1}{c}{97.50} & \multicolumn{1}{c}{97.90} & \multicolumn{1}{c}{98.20} & \multicolumn{1}{c}{98.80}\\

\multicolumn{1}{c}{} & \multicolumn{1}{c|}{PMS \cite{huang2012steganography}} & \multicolumn{1}{c}{/} & \multicolumn{1}{c}{/} & \multicolumn{1}{c}{/} & \multicolumn{1}{c}{/} & \multicolumn{1}{c}{/} & \multicolumn{1}{c}{/} & \multicolumn{1}{c}{/} & \multicolumn{1}{c}{/} & \multicolumn{1}{c}{/} & \multicolumn{1}{c}{/}\\
\hline
\multicolumn{1}{c}{\multirow{2}{*}{LStegT \cite{10574375}$^\dagger$}} & \multicolumn{1}{c|}{CNV-QIM \cite{xiao2008approach}} & \multicolumn{1}{c}{93.36} & \multicolumn{1}{c}{97.93} & \multicolumn{1}{c}{99.19} & \multicolumn{1}{c}{99.48} & \multicolumn{1}{c}{99.74} & \multicolumn{1}{c}{99.79} & \multicolumn{1}{c}{99.93} & \multicolumn{1}{c}{99.94} & \multicolumn{1}{c}{99.96} & \multicolumn{1}{c}{99.98}\\

\multicolumn{1}{c}{} & \multicolumn{1}{c|}{PMS \cite{huang2012steganography}} & \multicolumn{1}{c}{/} & \multicolumn{1}{c}{/} & \multicolumn{1}{c}{/} & \multicolumn{1}{c}{/} & \multicolumn{1}{c}{/} & \multicolumn{1}{c}{/} & \multicolumn{1}{c}{/} & \multicolumn{1}{c}{/} & \multicolumn{1}{c}{/} & \multicolumn{1}{c}{/}\\
\hline
\multicolumn{1}{c}{\multirow{2}{*}{FS-MDP \cite{2023Frame}$^\dagger$}} & \multicolumn{1}{c|}{CNV-QIM \cite{xiao2008approach}} & \multicolumn{1}{c}{69.76} & \multicolumn{1}{c}{74.74} & \multicolumn{1}{c}{77.17} & \multicolumn{1}{c}{82.27} & \multicolumn{1}{c}{88.44} & \multicolumn{1}{c}{88.11} & \multicolumn{1}{c}{84.95} & \multicolumn{1}{c}{86.11} & \multicolumn{1}{c}{90.78} & \multicolumn{1}{c}{88.69}\\

\multicolumn{1}{c}{} & \multicolumn{1}{c|}{PMS \cite{huang2012steganography}} & \multicolumn{1}{c}{/} & \multicolumn{1}{c}{/} & \multicolumn{1}{c}{/} & \multicolumn{1}{c}{/} & \multicolumn{1}{c}{/} & \multicolumn{1}{c}{/} & \multicolumn{1}{c}{/} & \multicolumn{1}{c}{/} & \multicolumn{1}{c}{/} & \multicolumn{1}{c}{/}\\
\hline
\multicolumn{1}{c}{\multirow{2}{*}{SFFN \cite{hu2021detection}}} & \multicolumn{1}{c|}{CNV-QIM \cite{xiao2008approach}} & \multicolumn{1}{c}{67.29} & \multicolumn{1}{c}{75.08} & \multicolumn{1}{c}{80.31} & \multicolumn{1}{c}{83.51} & \multicolumn{1}{c}{85.17} & \multicolumn{1}{c}{88.98} & \multicolumn{1}{c}{90.70} & \multicolumn{1}{c}{91.28} & \multicolumn{1}{c}{92.51} & \multicolumn{1}{c}{93.50}\\

\multicolumn{1}{c}{} & \multicolumn{1}{c|}{PMS \cite{huang2012steganography}} & \multicolumn{1}{c}{62.74} & \multicolumn{1}{c}{67.22} & \multicolumn{1}{c}{69.76} & \multicolumn{1}{c}{72.84} & \multicolumn{1}{c}{74.93} & \multicolumn{1}{c}{76.29} & \multicolumn{1}{c}{77.27} & \multicolumn{1}{c}{79.02} & \multicolumn{1}{c}{80.22} & \multicolumn{1}{c}{83.93}\\
\hline
\multicolumn{1}{c}{\multirow{2}{*}{KFEF \cite{wang2021fast}}} & \multicolumn{1}{c|}{CNV-QIM \cite{xiao2008approach}} & \multicolumn{1}{c}{79.59} & \multicolumn{1}{c}{87.87} & \multicolumn{1}{c}{91.98} & \multicolumn{1}{c}{93.89} & \multicolumn{1}{c}{95.70} & \multicolumn{1}{c}{96.81} & \multicolumn{1}{c}{97.57} & \multicolumn{1}{c}{98.12} & \multicolumn{1}{c}{98.46} & \multicolumn{1}{c}{98.90}\\

\multicolumn{1}{c}{} & \multicolumn{1}{c|}{PMS \cite{huang2012steganography}} & \multicolumn{1}{c}{62.87} & \multicolumn{1}{c}{67.23} & \multicolumn{1}{c}{70.39} & \multicolumn{1}{c}{74.05} & \multicolumn{1}{c}{76.42} & \multicolumn{1}{c}{77.48} & \multicolumn{1}{c}{79.41} & \multicolumn{1}{c}{81.05} & \multicolumn{1}{c}{81.92} & \multicolumn{1}{c}{85.76}\\
\hline
\hline
\multicolumn{1}{c}{\multirow{2}{*}{DVSF (ours)}} & \multicolumn{1}{c|}{CNV-QIM \cite{xiao2008approach}} & \multicolumn{1}{c}{\textbf{96.78}} & \multicolumn{1}{c}{\textbf{99.14}} & \multicolumn{1}{c}{\textbf{99.73}} & \multicolumn{1}{c}{\textbf{99.85}} & \multicolumn{1}{c}{\textbf{99.98}} & \multicolumn{1}{c}{\textbf{99.99}} & \multicolumn{1}{c}{\textbf{99.99}} & \multicolumn{1}{c}{\textbf{99.99}} & \multicolumn{1}{c}{\textbf{99.99}} & \multicolumn{1}{c}{\textbf{99.99}}\\

\multicolumn{1}{c}{} & \multicolumn{1}{c|}{PMS \cite{huang2012steganography}} & \multicolumn{1}{c}{\textbf{87.18}} & \multicolumn{1}{c}{\textbf{90.55}} & \multicolumn{1}{c}{\textbf{92.41}} & \multicolumn{1}{c}{\textbf{93.86}} & \multicolumn{1}{c}{\textbf{95.07}} & \multicolumn{1}{c}{\textbf{96.25}} & \multicolumn{1}{c}{\textbf{96.91}} & \multicolumn{1}{c}{\textbf{97.56}} & \multicolumn{1}{c}{\textbf{97.59}} & \multicolumn{1}{c}{\textbf{98.31}}\\

\midrule[1.5pt]

\end{tabularx}
\begin{threeparttable}
 \begin{tablenotes}
        \scriptsize
        \item[$\dagger$] We implement the methods on our datasets.
\end{tablenotes}
\end{threeparttable}
\end{table*}

In the architecture of our method, the overall joint loss function \( \mathcal{L}_{joint} \) consists of two components: the supervised contrastive learning training objective \( \mathcal{L}_{scl} \), which is designed to appropriately adjust the feature distances between normal and steganographic VoIP segments and enhance the model's sensitivity to hard-to-detect VoIP segments, and the cross-entropy loss training objective \( \mathcal{L}_{ce} \), which aims to accelerate model convergence and further improve its performance:
\begin{equation}
\mathcal{L}_{joint} = \mathcal{L}_{scl} + \mathcal{L}_{ce}.
\end{equation}

\section{Experiments and Analysis\protect\footnote{The data and code used in this paper are available at \href{https://github.com/PKQZPC/DVSF}{\color{black}https://github.com/PKQZPC/DVSF}.}}
In this section, we conducted several experiments. First, we introduce the dataset used and describe how it was expanded for more comprehensive experiments. Following that, we discuss the experimental setup and hyperparameters. We compare the proposed method with several of the best existing VoIP steganalysis methods and perform ablation studies to determine the contributions of key components in our approach. Additionally, we carry out visualization analysis and efficiency testing. Ultimately, we verify the effectiveness of the proposed VoIP steganalysis method from multiple perspectives.

\subsection{Experiment Setup}
To facilitate comparison with other methods, we use the state-of-the-art VStego800K dataset \cite{xu2023vstego800k}, which includes both CNV-QIM \cite{xiao2008approach} and PMS \cite{huang2012steganography} steganographic methods, as our base dataset. This dataset contains 814,592 VoIP stream segments, each with a duration of 1 second, and the embedding rates for steganographic samples range from 10\% to 40\%. To comprehensively showcase the advantages of our proposed method, we augmented this foundational dataset. Specifically, we expanded the range of embedding rates for the steganographic dataset to 10\% to 100\%, creating a new dataset \( D_e \) to demonstrate the superiority of our method over a broader range. To analyze the performance of our method in VoIP steganalysis for different segment lengths, we constructed a new dataset \( D_s \) by varying the segment length of VoIP streams from 0.1 s to 1.0 s, with all VoIP stream segments set to an embedding rate of 100\%. We particularly focus on the case of hard-to-detect VoIP segments with an embedding rate of 10\% and a segment length of 0.1 s.

To evaluate the performance of our proposed VoIP steganalysis architecture \(\textbf{DVSF}\) specifically designed for hard-to-detect samples, we select accuracy as the evaluation metric. This metric is intended to measure the model's ability to classify normal and steganographic VoIP segments. It is calculated as follows: 
\begin{equation}
Accuracy = \frac{TP+TN}{TP+FN+FP+FN},
\end{equation}
where TP denotes the count of positive samples correctly predicted as positive, FP indicates the number of negative samples that are incorrectly predicted as positive. FN represents the positive samples wrongly predicted as negative, while TN accounts for the negative samples accurately predicted as negative.

We use detection time to evaluate the processing speed of each method. Detection time is defined as the average time taken to process a fixed length of VoIP segments (e.g., 1 second). For a VoIP steganalysis method, if its detection time is less than the length of the VoIP segment itself, it meets the requirements for near-real-time detection.

The proposed architecture \(\textbf{DVSF}\) is built using PyTorch and accelerated using a Tesla V100-PCIE-32GB GPU. In our experiments, we set the learning rate of the architecture to 0.001, the batch size to 256, the temperature hyperparameter \(\tau\) to 0.025, and the number of training epochs to 300. In our dual-view feature extractor, the number of \textbf{HAM} blocks is set to 12.

\subsection{Comparison with the Existing Methods}
We compare the method proposed in this paper with several state-of-the-art VoIP steganalysis methods, including CCN \cite{li2014detection}, SS-QCCN \cite{li2017steganalysis}, LStegT \cite{10574375}, FS-MDP \cite{2023Frame}, SFFN \cite{hu2021detection}, and KFEF \cite{wang2021fast}. The experiments involve performance comparisons under different embedding rates and varying segment lengths. We focus on the detection results for challenging VoIP steganographic samples, particularly in scenarios with low embedding rates and short durations.

\begin{figure*}[t!]
\centering
\includegraphics[width=\textwidth]{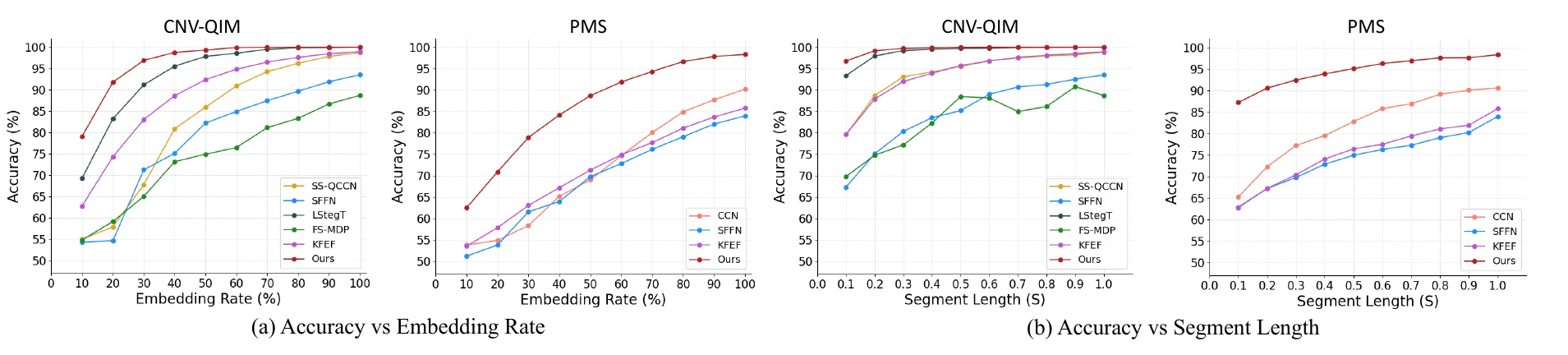}
\caption{For VoIP segment datasets \( D_e \) and \( D_s \), the detection accuracy varies with the embedding rate and segment length.}
\label{fig4}
\end{figure*}

\begin{table*}[h]
\centering
\caption{Average accuracy of DVSF on $D_e$ after eliminating or retaining only certain components.}
\label{table3}
\newcolumntype{Y}{>{\centering\arraybackslash}X}
\setlength{\extrarowheight}{2.5pt}
\begin{tabularx}{\textwidth}{YYYYYYYYYYYY}

\midrule[1.5pt]
\multicolumn{1}{l}{\textbf{Steganalysis}} & \multicolumn{1}{l|}{\textbf{Steganography}} & \multicolumn{10}{c}{\textbf{Embedding Rate (\%)}}\\

\multicolumn{1}{l}{\textbf{Method}} & \multicolumn{1}{l|}{\textbf{Method}} & \multicolumn{1}{Y}{10} & \multicolumn{1}{Y}{20} & \multicolumn{1}{Y}{30} & \multicolumn{1}{Y}{40} & \multicolumn{1}{Y}{50} & \multicolumn{1}{Y}{60} & \multicolumn{1}{Y}{70} & \multicolumn{1}{Y}{80} & \multicolumn{1}{Y}{90} & \multicolumn{1}{Y}{100}\\
\hline
\multicolumn{1}{c}{HAM} & \multicolumn{1}{c|}{CNV-QIM \cite{xiao2008approach}} & \multicolumn{1}{c}{73.38} & \multicolumn{1}{c}{87.18} & \multicolumn{1}{c}{94.20} & \multicolumn{1}{c}{96.98} & \multicolumn{1}{c}{98.73} & \multicolumn{1}{c}{99.39} & \multicolumn{1}{c}{99.62} & \multicolumn{1}{c}{99.91} & \multicolumn{1}{c}{99.98} & \multicolumn{1}{c}{99.99}\\

\multicolumn{1}{c}{(part of ours)} & \multicolumn{1}{c|}{PMS \cite{huang2012steganography}} & \multicolumn{1}{c}{59.55} & \multicolumn{1}{c}{68.54} & \multicolumn{1}{c}{76.43} & \multicolumn{1}{c}{82.52} & \multicolumn{1}{c}{87.01} & \multicolumn{1}{c}{90.14} & \multicolumn{1}{c}{93.04} & \multicolumn{1}{c}{95.35} & \multicolumn{1}{c}{97.25} & \multicolumn{1}{c}{97.99}\\
\hline
\multicolumn{1}{c}{HAM+JTS} & \multicolumn{1}{c|}{CNV-QIM \cite{xiao2008approach}} & \multicolumn{1}{c}{76.98} & \multicolumn{1}{c}{89.06} & \multicolumn{1}{c}{95.40} & \multicolumn{1}{c}{97.25} & \multicolumn{1}{c}{99.23} & \multicolumn{1}{c}{99.79} & \multicolumn{1}{c}{99.91} & \multicolumn{1}{c}{99.98} & \multicolumn{1}{c}{99.99} & \multicolumn{1}{c}{99.99}\\

\multicolumn{1}{c}{(part of ours)} & \multicolumn{1}{c|}{PMS \cite{huang2012steganography}} & \multicolumn{1}{c}{62.07} & \multicolumn{1}{c}{70.38} & \multicolumn{1}{c}{78.50} & \multicolumn{1}{c}{84.08} & \multicolumn{1}{c}{88.40} & \multicolumn{1}{c}{91.84} & \multicolumn{1}{c}{94.18} & \multicolumn{1}{c}{96.52} & \multicolumn{1}{c}{97.69} & \multicolumn{1}{c}{98.24}\\
\hline
\multicolumn{1}{c}{HAM+CutMix} & \multicolumn{1}{c|}{CNV-QIM \cite{xiao2008approach}} & \multicolumn{1}{c}{73.63} & \multicolumn{1}{c}{87.57} & \multicolumn{1}{c}{94.50} & \multicolumn{1}{c}{97.45} & \multicolumn{1}{c}{98.86} & \multicolumn{1}{c}{99.46} & \multicolumn{1}{c}{99.79} & \multicolumn{1}{c}{99.95} & \multicolumn{1}{c}{99.99} & \multicolumn{1}{c}{99.99}\\

\multicolumn{1}{c}{(part of ours)} & \multicolumn{1}{c|}{PMS \cite{huang2012steganography}} & \multicolumn{1}{c}{60.30} & \multicolumn{1}{c}{69.77} & \multicolumn{1}{c}{77.38} & \multicolumn{1}{c}{84.06} & \multicolumn{1}{c}{87.90} & \multicolumn{1}{c}{90.94} & \multicolumn{1}{c}{93.48} & \multicolumn{1}{c}{96.06} & \multicolumn{1}{c}{97.60} & \multicolumn{1}{c}{98.22}\\
\hline
\hline
\multicolumn{1}{c}{\multirow{2}{*}{DVSF (ours)}} & \multicolumn{1}{c|}{CNV-QIM \cite{xiao2008approach}} & \multicolumn{1}{c}{\textbf{79.03}} & \multicolumn{1}{c}{\textbf{91.75}} & \multicolumn{1}{c}{\textbf{96.88}} & \multicolumn{1}{c}{\textbf{98.72}} & \multicolumn{1}{c}{\textbf{99.27}} & \multicolumn{1}{c}{\textbf{99.89}} & \multicolumn{1}{c}{\textbf{99.95}} & \multicolumn{1}{c}{\textbf{99.99}} & \multicolumn{1}{c}{\textbf{99.99}} & \multicolumn{1}{c}{\textbf{99.99}}\\

\multicolumn{1}{c}{} & \multicolumn{1}{c|}{PMS \cite{huang2012steganography}} & \multicolumn{1}{c}{\textbf{62.45}} & \multicolumn{1}{c}{\textbf{70.85}} & \multicolumn{1}{c}{\textbf{78.90}} & \multicolumn{1}{c}{\textbf{84.15}} & \multicolumn{1}{c}{\textbf{88.66}} & \multicolumn{1}{c}{\textbf{91.86}} & \multicolumn{1}{c}{\textbf{94.32}} & \multicolumn{1}{c}{\textbf{96.64}} & \multicolumn{1}{c}{\textbf{97.80}} & \multicolumn{1}{c}{\textbf{98.31}}\\

\midrule[1.5pt]

\end{tabularx}
\end{table*}

\subsubsection{Performance Analysis at Different Embedding Rates.}
We conducted this experiment on the \(D_e\) dataset, and the results are presented in Table 1 and Figure 4 (a). Overall, for both steganographic methods, our approach outperforms the other four methods across various embedding rates, particularly showing a significant lead at lower embedding rates (below or equal to 50\%). For the more hard-to-detect VoIP segments with an embedding rate of 10\%, our method also demonstrates superior detection performance for the CNV-QIM steganographic method, achieving an accuracy of 79.03\%, which is nearly 10 percentage points higher than the current methods. For the PMS steganographic method, our method (62.45\%) considerably surpasses the best existing method, CCN (53.75\%). In practical application scenarios, our method exhibits significant advantages.

\subsubsection{Performance Analysis at Various VoIP Segment Lengths.}
We conducted this experiment on the \(D_s\) dataset, and the results are shown in Table 2 and Figure 4 (b). For both CNV-QIM and PMS steganographic methods, our approach demonstrated superior steganalysis performance when detecting VoIP segments of various lengths. Notably, for the hard-to-detect steganographic samples of these two methods with a segment duration of 0.1 seconds, our approach achieved remarkable improvements, with detection accuracies reaching 96.78\% and 87.18\%. This represents an improvement of 3.42\% and 21.93\% over the best-performing existing method, LStegT (93.36\%) and CCN (65.25\%), respectively. This indicates that our method demonstrates stronger adaptability in scenarios with shorter VoIP segments, significantly outperforming other methods.

\subsection{Ablation Analysis}
To further investigate the contribution of each component in our method to the overall performance and to explore the deeper relationships with the learning of hard-to-detect VoIP segments, we conducted three groups of ablation studies on the \(D_e\) dataset. The results are presented in Table 3. From this, we can observe that the specific Joint Training Strategy (\textbf{JTS}) significantly enhances the detection performance of the overall method, while the \textbf{HAM} model and the CutMix technique are also indispensable parts of the approach.

\subsubsection{Ablation Study of CutMix.}
From the experimental results of the two steganographic methods, the improvement brought by the CutMix technique to our approach is notable. For the CNV-QIM steganographic method, when the embedding rate is 10\%, the CutMix technique contributed to a 2.05\% increase in overall detection accuracy. For the PMS steganographic method, the improvement was 0.38\%. This enhancement is attributed to the fact that CutMix mixes the steganographic descriptors of VoIP segments, enabling the model to focus more on hard-to-detect VoIP segments.

\subsubsection{Ablation Study of Dual-View Feature Extractor.}
When our method consists solely of the dual-view feature extractor, it is evident that for the CNV-QIM steganographic method with an embedding rate of 10\%, the detection accuracy reaches 73.38\%, which exceeds that of the best existing method, KFEF, at 62.71\%. The same trend is observed for the PMS method, which also outperforms KFEF. This can be attributed to our model (\textbf{HAM}) being able to account for both the global and local features of VoIP segments, resulting in a richer feature space that encompasses more characteristics of hard-to-detect VoIP segments. Consequently, the model demonstrates greater generalization capability, leading to an improvement in detection accuracy.

\subsubsection{Ablation Study of Joint Training Strategy.}
From the experimental results, we can see that replacing the existing cross-entropy loss learning strategy with our Joint Training Strategy (\textbf{JTS}) leads to a significant improvement in the detection accuracy for the CNV-QIM steganographic method at an embedding rate of 10\%, achieving an increase of 5.40\%. Even for the more hard-to-detect PMS steganographic method, our approach still realizes a 2.15\% improvement in detection accuracy. This improvement is attributed to the construction of closely-related cyclic triplets during the training process, which encourages the features of normal and steganographic VoIP to become linearly separable. This approach optimizes the alignment and uniformity \cite{wang2020understanding} of the model's feature space, enabling the model to achieve better fitting for the features of hard-to-detect VoIP segments.

\begin{figure*}[t]
\centering
\includegraphics[width=\textwidth]{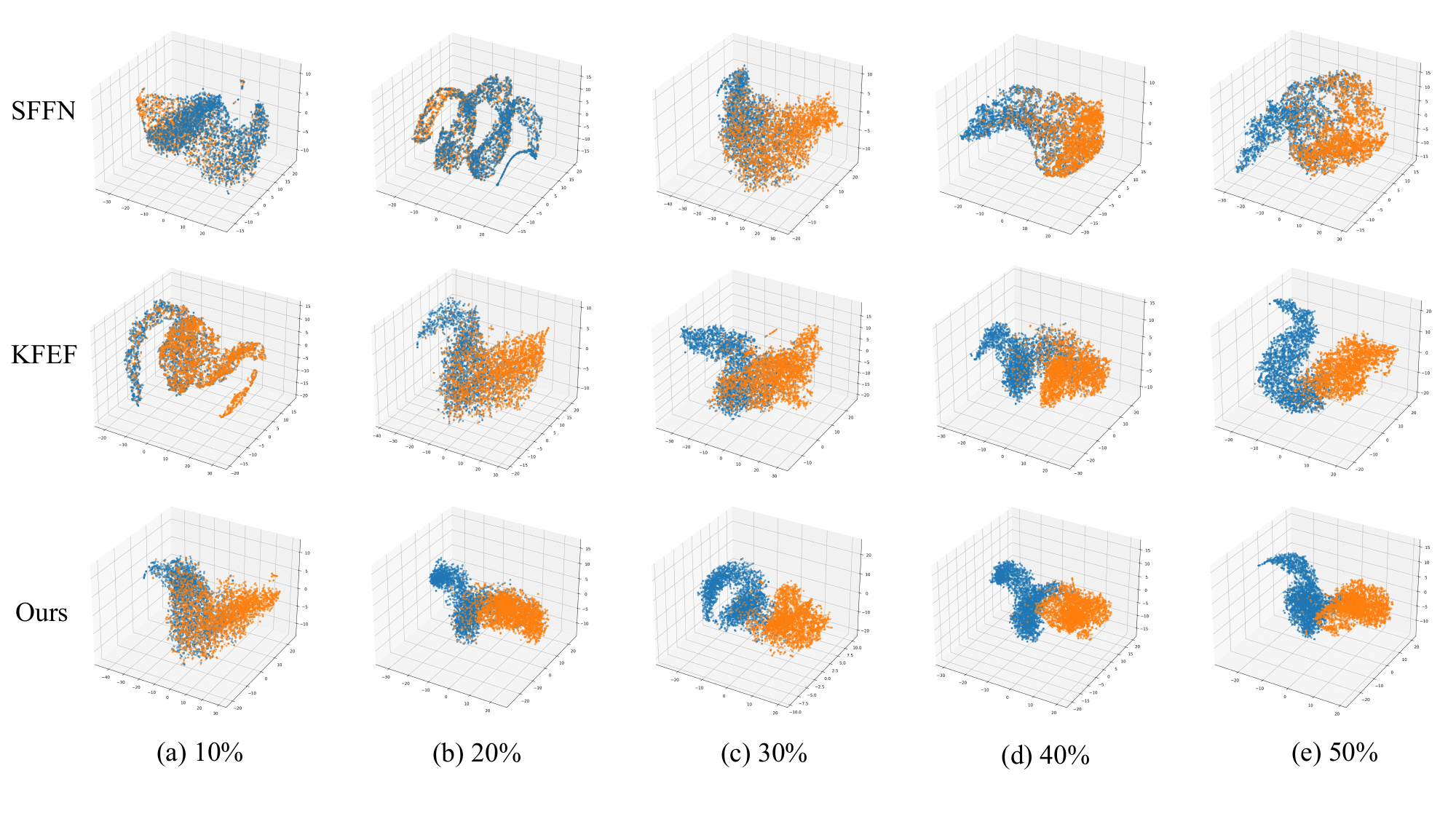}
\caption{The distribution of speech features extracted by the proposed method and the best existing methods in statistical space varies with the embedding rate. Each of the dots represents a speech with a length of 1s, the orange points indicate cover speeches, and the blue points indicate stego speeches with different embedding rate concealment information.}
\label{fig4}
\end{figure*}

\subsection{Visualization Analysis}
To visually demonstrate the superiority of our method, we employed t-distributed Stochastic Neighbor Embedding (t-SNE) \cite{van2014accelerating} to reduce the dimensionality of the features obtained from the three VoIP steganalysis methods in the experiment for steganographic samples with embedding rates ranging from 10\% to 50\%, as shown in Figure 5. As the embedding rate increases, we observe a gradual separation between normal VoIP and steganographic VoIP in the feature space. At the same embedding rate, particularly in the hard-to-detect scenario with an embedding rate of 10\%, our method clearly exhibits superior classification performance in the feature space compared to the other two methods. The results in Figure 5 intuitively reflect the proposed model's stronger ability to extract and analyze VoIP steganographic features.

\begin{table}[h]
\centering
\caption{Average detection time for a 1-second VoIP segment.}
\label{table1}
\setlength{\extrarowheight}{2pt}
\begin{tabularx}{0.85\columnwidth}{llllllllc}

\midrule[1.5pt]
\multicolumn{1}{l}{\textbf{Steganalysis Method}} & & & & & & & & \textbf{Time (ms)}\\
\hline
\multicolumn{1}{l}{CCN \cite{li2014detection}} & & & & & & & & 0.021\\
\multicolumn{1}{l}{SS-QCCN \cite{li2017steganalysis}} & & & & & & & & 0.284\\
\multicolumn{1}{l}{LStegT \cite{10574375}$^\dagger$} & & & & & & & & 0.607\\
\multicolumn{1}{l}{FS-MDP \cite{2023Frame}$^\dagger$} & & & & & & & & 0.681\\
\multicolumn{1}{l}{SFFN \cite{hu2021detection}} & & & & & & & & 0.223\\
\multicolumn{1}{l}{KFEF \cite{wang2021fast}} & & & & & & & & 0.088\\
\hline
\multicolumn{1}{l}{DVSF (ours)} & & & & & & & & 0.010\\
\midrule[1.5pt]
\end{tabularx}
\begin{threeparttable}
 \begin{tablenotes}
        \scriptsize
        \item[$\dagger$] We implement the methods on our datasets.
\end{tablenotes}
\end{threeparttable}
\end{table}

\subsection{Time Efficiency Testing}
Since VoIP audio signals are transmitted in real-time online, VoIP steganalysis methods must meet near-real-time requirement. We tested all methods from the experimental section of this paper, with results shown in Table 4. From the results, we can see that our method detects a 1-second VoIP segment in the shortest time, operating at twice the speed of the currently fastest CCN \cite{li2014detection}. Since FS-MDP \cite{2023Frame} is a frame-level detection method, it requires more time for frame processing, making it the slowest with a speed of 0.681 ms. Overall, our method requires only 0.010 ms to detect a 1-second VoIP segment, meeting the demand for near-real-time steganalysis and outperforming the current best methods.

\section{Conclusion}
In this paper, we propose a Dual-View VoIP Steganalysis Framework (\textbf{DVSF}) and use CutMix technology to adjust VoIP segment features. The core of the dual-view feature extractor is Hybrid Attention Model (\textbf{HAM}) that focuses on the fusion analysis of local and global steganographic features in VoIP. Additionally, we utilize a Joint Training Strategy (\textbf{JTS}) based on supervised contrastive learning with specially constructed VoIP segment triplets for model optimization. Extensive experiments demonstrate that our method effectively overcomes the challenge of detecting hard-to-detect VoIP segments and significantly outperforms the current state-of-the-art detection methods across various steganalysis scenarios, including those with low embedding rates and short durations. Additionally, our method achieves the best detection efficiency, making it better suited to meet the demands of near-real-time detection. We hope this work will accelerate the development of the VoIP steganalysis field, providing more practical significance.

\bibliographystyle{plain}
\bibliography{ref}

\end{document}